\documentclass[conference,10pt,twocolumn,letter]{IEEEtran}
\IEEEoverridecommandlockouts
\usepackage{cite}
\usepackage[T1]{fontenc}
\usepackage{graphicx}
\usepackage{amssymb}
\usepackage{amsmath}
\usepackage{paralist}
\usepackage{subfigure}
\usepackage{booktabs} 
\usepackage{algorithm}
\usepackage{algorithmic}
\usepackage{amsthm}
\usepackage{multirow}
\usepackage{microtype} 
\usepackage{tablefootnote}
\usepackage{balance}

\usepackage{amssymb}
\usepackage{amsfonts}
\usepackage{mathrsfs}
\usepackage{xspace}
\usepackage{bm}
\usepackage{upgreek}

\newcommand{\safemath}[2]{\newcommand{#1}{\ensuremath{#2}\xspace}}

\safemath{\bma}{\mathbf{a}}
\safemath{\bmb}{\mathbf{b}}
\safemath{\bmc}{\mathbf{c}}
\safemath{\bmd}{\mathbf{d}}
\safemath{\bme}{\mathbf{e}}
\safemath{\bmf}{\mathbf{f}}
\safemath{\bmg}{\mathbf{g}}
\safemath{\bmh}{\mathbf{h}}
\safemath{\bmi}{\mathbf{i}}
\safemath{\bmj}{\mathbf{j}}
\safemath{\bmk}{\mathbf{k}}
\safemath{\bml}{\mathbf{l}}
\safemath{\bmm}{\mathbf{m}}
\safemath{\bmn}{\mathbf{n}}
\safemath{\bmo}{\mathbf{o}}
\safemath{\bmp}{\mathbf{p}}
\safemath{\bmq}{\mathbf{q}}
\safemath{\bmr}{\mathbf{r}}
\safemath{\bms}{\mathbf{s}}
\safemath{\bmt}{\mathbf{t}}
\safemath{\bmu}{\mathbf{u}}
\safemath{\bmv}{\mathbf{v}}
\safemath{\bmw}{\mathbf{w}}
\safemath{\bmx}{\mathbf{x}}
\safemath{\bmy}{\mathbf{y}}
\safemath{\bmz}{\mathbf{z}}
\safemath{\bmzero}{\mathbf{0}}
\safemath{\bmone}{\mathbf{1}}

\bmdefine{\biad}{a}
\bmdefine{\bibd}{b}
\bmdefine{\bicd}{c}
\bmdefine{\bidd}{d}
\bmdefine{\bied}{e}
\bmdefine{\bifd}{f}
\bmdefine{\bigd}{g}
\bmdefine{\bihd}{h}
\bmdefine{\biid}{i}
\bmdefine{\bijd}{j}
\bmdefine{\bikd}{k}
\bmdefine{\bild}{l}
\bmdefine{\bimd}{m}
\bmdefine{\bind}{n}
\bmdefine{\biod}{o}
\bmdefine{\bipd}{p}
\bmdefine{\biqd}{q}
\bmdefine{\bird}{r}
\bmdefine{\bisd}{s}
\bmdefine{\bitd}{t}
\bmdefine{\biud}{u}
\bmdefine{\bivd}{v}
\bmdefine{\biwd}{w}
\bmdefine{\bixd}{x}
\bmdefine{\biyd}{y}
\bmdefine{\bizd}{z}

\bmdefine{\bixid}{\xi}
\bmdefine{\bilambdad}{\lambda}
\bmdefine{\bimud}{\mu}
\bmdefine{\bithetad}{\theta}
\bmdefine{\biphid}{\phi}
\bmdefine{\bideltad}{\delta}

\safemath{\bmia}{\biad}
\safemath{\bmib}{\bibd}
\safemath{\bmic}{\bicd}
\safemath{\bmid}{\bidd}
\safemath{\bmie}{\bied}
\safemath{\bmif}{\bifd}
\safemath{\bmig}{\bigd}
\safemath{\bmih}{\bihd}
\safemath{\bmii}{\biid}
\safemath{\bmij}{\bijd}
\safemath{\bmik}{\bikd}
\safemath{\bmil}{\bild}
\safemath{\bmim}{\bimd}
\safemath{\bmin}{\bind}
\safemath{\bmio}{\biod}
\safemath{\bmip}{\bipd}
\safemath{\bmiq}{\biqd}
\safemath{\bmir}{\bird}
\safemath{\bmis}{\bisd}
\safemath{\bmit}{\bitd}
\safemath{\bmiu}{\biud}
\safemath{\bmiv}{\bivd}
\safemath{\bmiw}{\biwd}
\safemath{\bmix}{\bixd}
\safemath{\bmiy}{\biyd}
\safemath{\bmiz}{\bizd}

\safemath{\bmxi}{\bixid}
\safemath{\bmlambda}{\bilambdad}
\safemath{\bmmu}{\bimud}
\safemath{\bmtheta}{\bithetad}
\safemath{\bmphi}{\biphid}
\safemath{\bmdelta}{\bideltad}

\safemath{\bA}{\mathbf{A}}
\safemath{\bB}{\mathbf{B}}
\safemath{\bC}{\mathbf{C}}
\safemath{\bD}{\mathbf{D}}
\safemath{\bE}{\mathbf{E}}
\safemath{\bF}{\mathbf{F}}
\safemath{\bG}{\mathbf{G}}
\safemath{\bH}{\mathbf{H}}
\safemath{\bI}{\mathbf{I}}
\safemath{\bJ}{\mathbf{J}}
\safemath{\bK}{\mathbf{K}}
\safemath{\bL}{\mathbf{L}}
\safemath{\bM}{\mathbf{M}}
\safemath{\bN}{\mathbf{N}}
\safemath{\bO}{\mathbf{O}}
\safemath{\bP}{\mathbf{P}}
\safemath{\bQ}{\mathbf{Q}}
\safemath{\bR}{\mathbf{R}}
\safemath{\bS}{\mathbf{S}}
\safemath{\bT}{\mathbf{T}}
\safemath{\bU}{\mathbf{U}}
\safemath{\bV}{\mathbf{V}}
\safemath{\bW}{\mathbf{W}}
\safemath{\bX}{\mathbf{X}}
\safemath{\bY}{\mathbf{Y}}
\safemath{\bZ}{\mathbf{Z}}

\safemath{\bZero}{\mathbf{0}}
\safemath{\bOne}{\mathbf{1}}
\safemath{\bDelta}{\mathbf{\Delta}}
\safemath{\bLambda}{\mathbf{\UpLambda}}
\safemath{\bPhi}{\mathbf{\Upphi}}
\safemath{\bSigma}{\mathbf{\Upsigma}}
\safemath{\bOmega}{\mathbf{\Upomega}}
\safemath{\bTheta}{\mathbf{\Uptheta}}

\bmdefine{\biAd}{A}
\bmdefine{\biBd}{B}
\bmdefine{\biCd}{C}
\bmdefine{\biDd}{D}
\bmdefine{\biEd}{E}
\bmdefine{\biFd}{F}
\bmdefine{\biGd}{G}
\bmdefine{\biHd}{H}
\bmdefine{\biId}{I}
\bmdefine{\biJd}{J}
\bmdefine{\biKd}{K}
\bmdefine{\biLd}{L}
\bmdefine{\biMd}{M}
\bmdefine{\biOd}{N}
\bmdefine{\biPd}{O}
\bmdefine{\biQd}{P}
\bmdefine{\biRd}{R}
\bmdefine{\biSd}{S}
\bmdefine{\biTd}{T}
\bmdefine{\biUd}{U}
\bmdefine{\biVd}{V}
\bmdefine{\biWd}{W}
\bmdefine{\biXd}{X}
\bmdefine{\biYd}{Y}
\bmdefine{\biZd}{Z}

\bmdefine{\biDelta}{\Delta}
\bmdefine{\biLambda}{\Lambda}
\bmdefine{\biPhi}{\Phi}
\bmdefine{\biSigma}{\Sigma}
\bmdefine{\biOmega}{\Omega}
\bmdefine{\biTheta}{\Theta}

\safemath{\bimA}{\biAd}
\safemath{\bimB}{\biBd}
\safemath{\bimC}{\biCd}
\safemath{\bimD}{\biDd}
\safemath{\bimE}{\biEd}
\safemath{\bimF}{\biFd}
\safemath{\bimG}{\biGd}
\safemath{\bimH}{\biHd}
\safemath{\bimI}{\biId}
\safemath{\bimJ}{\biJd}
\safemath{\bimK}{\biKd}
\safemath{\bimL}{\biLd}
\safemath{\bimM}{\biMd}
\safemath{\bimN}{\biNd}
\safemath{\bimO}{\biOd}
\safemath{\bimP}{\biPd}
\safemath{\bimQ}{\biQd}
\safemath{\bimR}{\biRd}
\safemath{\bimS}{\biSd}
\safemath{\bimT}{\biTd}
\safemath{\bimU}{\biUd}
\safemath{\bimV}{\biVd}
\safemath{\bimW}{\biWd}
\safemath{\bimX}{\biXd}
\safemath{\bimY}{\biYd}
\safemath{\bimZ}{\biZd}

\safemath{\bimDelta}{\biDelta}
\safemath{\bimLambda}{\biLambda}
\safemath{\bimPhi}{\biPhi}
\safemath{\bimSigma}{\biSigma}
\safemath{\bimOmega}{\biOmega}
\safemath{\bimTheta}{\biTheta}

\safemath{\setA}{\mathcal{A}}
\safemath{\setB}{\mathcal{B}}
\safemath{\setC}{\mathcal{C}}
\safemath{\setD}{\mathcal{D}}
\safemath{\setE}{\mathcal{E}}
\safemath{\setF}{\mathcal{F}}
\safemath{\setG}{\mathcal{G}}
\safemath{\setH}{\mathcal{H}}
\safemath{\setI}{\mathcal{I}}
\safemath{\setJ}{\mathcal{J}}
\safemath{\setK}{\mathcal{K}}
\safemath{\setL}{\mathcal{L}}
\safemath{\setM}{\mathcal{M}}
\safemath{\setN}{\mathcal{N}}
\safemath{\setO}{\mathcal{O}}
\safemath{\setP}{\mathcal{P}}
\safemath{\setQ}{\mathcal{Q}}
\safemath{\setR}{\mathcal{R}}
\safemath{\setS}{\mathcal{S}}
\safemath{\setT}{\mathcal{T}}
\safemath{\setU}{\mathcal{U}}
\safemath{\setV}{\mathcal{V}}
\safemath{\setW}{\mathcal{W}}
\safemath{\setX}{\mathcal{X}}
\safemath{\setY}{\mathcal{Y}}
\safemath{\setZ}{\mathcal{Z}}
\safemath{\emptySet}{\varnothing}

\safemath{\colA}{\mathscr{A}}
\safemath{\colB}{\mathscr{B}}
\safemath{\colC}{\mathscr{C}}
\safemath{\colD}{\mathscr{D}}
\safemath{\colE}{\mathscr{E}}
\safemath{\colF}{\mathscr{F}}
\safemath{\colG}{\mathscr{G}}
\safemath{\colH}{\mathscr{H}}
\safemath{\colI}{\mathscr{I}}
\safemath{\colJ}{\mathscr{J}}
\safemath{\colK}{\mathscr{K}}
\safemath{\colL}{\mathscr{L}}
\safemath{\colM}{\mathscr{M}}
\safemath{\colN}{\mathscr{N}}
\safemath{\colO}{\mathscr{O}}
\safemath{\colP}{\mathscr{P}}
\safemath{\colQ}{\mathscr{Q}}
\safemath{\colR}{\mathscr{R}}
\safemath{\colS}{\mathscr{S}}
\safemath{\colT}{\mathscr{T}}
\safemath{\colU}{\mathscr{U}}
\safemath{\colV}{\mathscr{V}}
\safemath{\colW}{\mathscr{W}}
\safemath{\colX}{\mathscr{X}}
\safemath{\colY}{\mathscr{Y}}
\safemath{\colZ}{\mathscr{Z}}

\safemath{\opA}{\mathbb{A}}
\safemath{\opB}{\mathbb{B}}
\safemath{\opC}{\mathbb{C}}
\safemath{\opD}{\mathbb{D}}
\safemath{\opE}{\mathbb{E}}
\safemath{\opF}{\mathbb{F}}
\safemath{\opG}{\mathbb{G}}
\safemath{\opH}{\mathbb{H}}
\safemath{\opI}{\mathbb{I}}
\safemath{\opJ}{\mathbb{J}}
\safemath{\opK}{\mathbb{K}}
\safemath{\opL}{\mathbb{L}}
\safemath{\opM}{\mathbb{M}}
\safemath{\opN}{\mathbb{N}}
\safemath{\opO}{\mathbb{O}}
\safemath{\opP}{\mathbb{P}}
\safemath{\opQ}{\mathbb{Q}}
\safemath{\opR}{\mathbb{R}}
\safemath{\opS}{\mathbb{S}}
\safemath{\opT}{\mathbb{T}}
\safemath{\opU}{\mathbb{U}}
\safemath{\opV}{\mathbb{V}}
\safemath{\opW}{\mathbb{W}}
\safemath{\opX}{\mathbb{X}}
\safemath{\opY}{\mathbb{Y}}
\safemath{\opZ}{\mathbb{Z}}
\safemath{\opZero}{\mathbb{O}}
\safemath{\identityop}{\opI}

\safemath{\veca}{\bma}
\safemath{\vecb}{\bmb}
\safemath{\vecc}{\bmc}
\safemath{\vecd}{\bmd}
\safemath{\vece}{\bme}
\safemath{\vecf}{\bmf}
\safemath{\vecg}{\bmg}
\safemath{\vech}{\bmh}
\safemath{\veci}{\bmi}
\safemath{\vecj}{\bmj}
\safemath{\veck}{\bmk}
\safemath{\vecl}{\bml}
\safemath{\vecm}{\bmm}
\safemath{\vecn}{\bmn}
\safemath{\veco}{\bmo}
\safemath{\vecp}{\bmp}
\safemath{\vecq}{\bmq}
\safemath{\vecr}{\bmr}
\safemath{\vecs}{\bms}
\safemath{\vect}{\bmt}
\safemath{\vecu}{\bmu}
\safemath{\vecv}{\bmv}
\safemath{\vecw}{\bmw}
\safemath{\vecx}{\bmx}
\safemath{\vecy}{\bmy}
\safemath{\vecz}{\bmz}

\safemath{\veczero}{\bmzero}
\safemath{\vecone}{\bmone}
\safemath{\vecxi}{\bmxi}
\safemath{\veclambda}{\bmlambda}
\safemath{\vecmu}{\bmmu}
\safemath{\vectheta}{\bmtheta}
\safemath{\vecphi}{\bmphi}
\safemath{\vecdelta}{\bmdelta}

\safemath{\matA}{\bA}
\safemath{\matB}{\bB}
\safemath{\matC}{\bC}
\safemath{\matD}{\bD}
\safemath{\matE}{\bE}
\safemath{\matF}{\bF}
\safemath{\matG}{\bG}
\safemath{\matH}{\bH}
\safemath{\matI}{\bI}
\safemath{\matJ}{\bJ}
\safemath{\matK}{\bK}
\safemath{\matL}{\bL}
\safemath{\matM}{\bM}
\safemath{\matN}{\bN}
\safemath{\matO}{\bO}
\safemath{\matP}{\bP}
\safemath{\matQ}{\bQ}
\safemath{\matR}{\bR}
\safemath{\matS}{\bS}
\safemath{\matT}{\bT}
\safemath{\matU}{\bU}
\safemath{\matV}{\bV}
\safemath{\matW}{\bW}
\safemath{\matX}{\bX}
\safemath{\matY}{\bY}
\safemath{\matZ}{\bZ}
\safemath{\matzero}{\bmzero}

\safemath{\matDelta}{\bDelta}
\safemath{\matLambda}{\bLambda}
\safemath{\matPhi}{\bPhi}
\safemath{\matSigma}{\bSigma}
\safemath{\matOmega}{\bOmega}
\safemath{\matTheta}{\bTheta}

\safemath{\matidentity}{\matI}
\safemath{\matone}{\matO}

\safemath{\rnda}{A}
\safemath{\rndb}{B}
\safemath{\rndc}{C}
\safemath{\rndd}{D}
\safemath{\rnde}{E}
\safemath{\rndf}{F}
\safemath{\rndg}{G}
\safemath{\rndh}{H}
\safemath{\rndi}{I}
\safemath{\rndj}{J}
\safemath{\rndk}{K}
\safemath{\rndl}{L}
\safemath{\rndm}{M}
\safemath{\rndn}{N}
\safemath{\rndo}{O}
\safemath{\rndp}{P}
\safemath{\rndq}{Q}
\safemath{\rndr}{R}
\safemath{\rnds}{S}
\safemath{\rndt}{T}
\safemath{\rndu}{U}
\safemath{\rndv}{V}
\safemath{\rndw}{W}
\safemath{\rndx}{X}
\safemath{\rndy}{Y}
\safemath{\rndz}{Z}

\safemath{\rveca}{\bimA}
\safemath{\rvecb}{\bimB}
\safemath{\rvecc}{\bimC}
\safemath{\rvecd}{\bimD}
\safemath{\rvece}{\bimE}
\safemath{\rvecf}{\bimF}
\safemath{\rvecg}{\bimG}
\safemath{\rvech}{\bimH}
\safemath{\rveci}{\bimI}
\safemath{\rvecj}{\bimJ}
\safemath{\rveck}{\bimK}
\safemath{\rvecl}{\bimL}
\safemath{\rvecm}{\bimM}
\safemath{\rvecn}{\bimN}
\safemath{\rveco}{\bomO}
\safemath{\rvecp}{\bimP}
\safemath{\rvecq}{\bimQ}
\safemath{\rvecr}{\bimR}
\safemath{\rvecs}{\bimS}
\safemath{\rvect}{\bimT}
\safemath{\rvecu}{\bimU}
\safemath{\rvecv}{\bimV}
\safemath{\rvecw}{\bimW}
\safemath{\rvecx}{\bimX}
\safemath{\rvecy}{\bimY}
\safemath{\rvecz}{\bimZ}

\safemath{\rvecxi}{\bmxi}
\safemath{\rveclambda}{\bmlambda}
\safemath{\rvecmu}{\bmmu}
\safemath{\rvectheta}{\bmtheta}
\safemath{\rvecphi}{\bmphi}

\safemath{\rmatA}{\bimA}
\safemath{\rmatB}{\bimB}
\safemath{\rmatC}{\bimC}
\safemath{\rmatD}{\bimD}
\safemath{\rmatE}{\bimE}
\safemath{\rmatF}{\bimF}
\safemath{\rmatG}{\bimG}
\safemath{\rmatH}{\bimH}
\safemath{\rmatI}{\bimI}
\safemath{\rmatJ}{\bimJ}
\safemath{\rmatK}{\bimK}
\safemath{\rmatL}{\bimL}
\safemath{\rmatM}{\bimM}
\safemath{\rmatN}{\bimN}
\safemath{\rmatO}{\bimO}
\safemath{\rmatP}{\bimP}
\safemath{\rmatQ}{\bimQ}
\safemath{\rmatR}{\bimR}
\safemath{\rmatS}{\bimS}
\safemath{\rmatT}{\bimT}
\safemath{\rmatU}{\bimU}
\safemath{\rmatV}{\bimV}
\safemath{\rmatW}{\bimW}
\safemath{\rmatX}{\bimX}
\safemath{\rmatY}{\bimY}
\safemath{\rmatZ}{\bimZ}

\safemath{\rmatDelta}{\bimDelta}
\safemath{\rmatLambda}{\bimLambda}
\safemath{\rmatPhi}{\bimPhi}
\safemath{\rmatSigma}{\bimSigma}
\safemath{\rmatOmega}{\bimOmega}
\safemath{\rmatTheta}{\bimTheta}

\usepackage{amssymb}
\usepackage{amsfonts}
\usepackage{mathrsfs}
\usepackage{xspace}
\usepackage{bm}
\usepackage{fancyref}
\usepackage{textcomp}

\usepackage{multirow}
\usepackage{stmaryrd}

\newenvironment{textbmatrix}{	\setlength{\arraycolsep}{2.5pt}%
								\big[\begin{matrix}}{\end{matrix}\big]%
								\raisebox{0.08ex}{\vphantom{M}}}

\def\be{\begin{equation}}
\def\ee{\end{equation}}
\def\een{\nonumber \end{equation}}
\def\mat{\begin{bmatrix}}
\def\emat{\end{bmatrix}}
\def\btm{\begin{textbmatrix}}
\def\etm{\end{textbmatrix}}

\def\ba#1\ea{\begin{align}#1\end{align}}
\def\bas#1\eas{\begin{align*}#1\end{align*}}
\def\bs#1\es{\begin{split}#1\end{split}}
\def\bg#1\eg{\begin{gather}#1\end{gather}}
\def\bml#1\eml{\begin{multline}#1\end{multline}}
\def\bi#1\ei{\begin{itemize}#1\end{itemize}}

\DeclareMathOperator{\sign}{sign}			

\safemath{\dirac}{\delta}					
\safemath{\krond}{\dirac}					

\safemath{\upto}{\uparrow}
\safemath{\downto}{\downarrow}
\safemath{\iu}{j}							
\safemath{\ev}{\lambda}						
\safemath{\hilseqspace}{l^{2}}				
\newcommand{\banachfunspace}[1]{\setL^{#1}}	
\safemath{\hilfunspace}{\banachfunspace{2}}	

\safemath{\SNR}{\textit{SNR}} 				
\safemath{\PAR}{\textit{PAR}} 				
\safemath{\No}{N_0}							
\safemath{\Es}{E_s}							
\safemath{\Eb}{E_b}							
\safemath{\EbNo}{\frac{\Eb}{\No}}
\safemath{\EsNo}{\frac{\Es}{\No}}

\DeclareMathOperator{\CHop}{\ensuremath{\opH}} 
\safemath{\tvir}{\rndh_{\CHop}}				
\safemath{\tvtf}{\rndl_{\CHop}}				
\safemath{\spf}{\rnds_{\CHop}}				
\safemath{\bff}{H_{\CHop}}					

\safemath{\ircf}{r_{h}}						
\safemath{\tftvcf}{r_{s}}					
\safemath{\tfcf}{r_{l}}						
\safemath{\bfcf}{r_{H}}						

\safemath{\tcorr}{c_h}						
\safemath{\scf}{c_{s}}						
\safemath{\tfcorr}{c_{l}}					
\safemath{\fcorr}{c_{H}}						

\safemath{\mi}{I}							
\safemath{\capacity}{C}						

\safemath{\normal}{\mathcal{N}}			
\safemath{\jpg}{\mathcal{CN}}			
\safemath{\mchain}{\leftrightarrow}		

\safemath{\dB}{\,\mathrm{dB}}
\safemath{\dBm}{\,\mathrm{dBm}}
\safemath{\Hz}{\,\mathrm{Hz}}
\safemath{\kHz}{\,\mathrm{kHz}}
\safemath{\MHz}{\,\mathrm{MHz}}
\safemath{\GHz}{\,\mathrm{GHz}}
\safemath{\s}{\,\mathrm{s}}
\safemath{\ms}{\,\mathrm{ms}}
\safemath{\mus}{\,\mathrm{\text{\textmu}s}}
\safemath{\ns}{\,\mathrm{ns}}
\safemath{\ps}{\,\mathrm{ps}}
\safemath{\meter}{\,\mathrm{m}}
\safemath{\mm}{\,\mathrm{mm}}
\safemath{\cm}{\,\mathrm{cm}}
\safemath{\m}{\,\mathrm{m}}
\safemath{\W}{\,\mathrm{W}}
\safemath{\mW}{\, \mathrm{mW}}
\safemath{\J}{\,\mathrm{J}}
\safemath{\K}{\,\mathrm{K}}
\safemath{\bit}{\,\mathrm{bit}}
\safemath{\nat}{\,\mathrm{nat}}

\safemath{\define}{\triangleq}			

\safemath{\equivalent}{\sim}
\safemath{\distas}{\sim}					
\safemath{\sdiff}{\Delta}				

\safemath{\reals}{\mathbb{R}}
\safemath{\positivereals}{\reals_{+}}
\safemath{\integers}{\mathbb{Z}}
\safemath{\posint}{\integers_{+}}
\safemath{\naturals}{\mathbb{N}}
\safemath{\posnaturals}{\naturals_{+}}
\safemath{\complexset}{\mathbb{C}}
\safemath{\rationals}{\mathbb{Q}}

\newcommand*{\fancyrefapplabelprefix}{app}		
\newcommand*{\fancyrefthmlabelprefix}{thm}		
\newcommand*{\fancyreflemlabelprefix}{lem}		
\newcommand*{\fancyrefcorlabelprefix}{cor}		
\newcommand*{\fancyrefdeflabelprefix}{def}		
\newcommand*{\fancyrefproplabelprefix}{prop}		
\newcommand*{\fancyrefexmpllabelprefix}{exmpl}
\newcommand*{\fancyrefalglabelprefix}{alg}		
\newcommand*{\fancyreftbllabelprefix}{tbl}		

\frefformat{vario}{\fancyrefseclabelprefix}{Section~#1}
\frefformat{vario}{\fancyrefthmlabelprefix}{Theorem~#1}
\frefformat{vario}{\fancyreftbllabelprefix}{Table~#1}
\frefformat{vario}{\fancyreflemlabelprefix}{Lemma~#1}
\frefformat{vario}{\fancyrefcorlabelprefix}{Corollary~#1}
\frefformat{vario}{\fancyrefdeflabelprefix}{Definition~#1}
\frefformat{vario}{\fancyreffiglabelprefix}{Fig.~#1}
\frefformat{vario}{\fancyrefapplabelprefix}{Appendix~#1}
\frefformat{vario}{\fancyrefeqlabelprefix}{(#1)}
\frefformat{vario}{\fancyrefproplabelprefix}{Proposition~#1}
\frefformat{vario}{\fancyrefexmpllabelprefix}{Example~#1}
\frefformat{vario}{\fancyrefalglabelprefix}{Algorithm~#1}

\safemath{\dictab}{[\,\dicta\,\,\dictb\,]}

\safemath{\ysig}{\bmy}
\safemath{\ysighat}{\hat{\ysig}}
\safemath{\ysigdim}{M}
\safemath{\xsig}{\bmx}
\safemath{\xsigdim}{N}
\safemath{\nx}{n_x}
\safemath{\zsig}{\bmz}
\safemath{\zsigdim}{\ysigdim}
\safemath{\rsig}{\bmr}
\safemath{\Adict}{\bA}
\safemath{\Adicttilde}{\widetilde{\Adict}}
\safemath{\Adictdim}{\outputdim\times\xsigdim}
\safemath{\avec}{\bma}
\safemath{\avectilde}{\tilde{\avec}}
\safemath{\Bdict}{\bB}
\safemath{\Bdicttilde}{\widetilde{\Bdict}}
\safemath{\Cdict}{\bC}
\safemath{\cvec}{\bmc}
\safemath{\Ddict}{\bD}
\safemath{\Ddictdim}{\ysigdim\times\xsigdim}
\safemath{\dvec}{\bmd}
\safemath{\Ddicttilde}{\widetilde{\bD}}
\safemath{\Bonb}{\bB}
\safemath{\bvec}{\bmb}
\safemath{\Bonbdim}{\ysigdim\times\ysigdim}
\safemath{\noise}{\bmn}
\safemath{\noisedim}{\ysigim}
\safemath{\err}{\bme}
\safemath{\errdim}{\ysigdim}
\safemath{\errset}{\setE}
\safemath{\nerr}{n_e}
\safemath{\delop}{\bP_\errset}
\safemath{\delopc}{\bP_{{\errset}^c}}

\safemath{\cplxi}{\imath}
\safemath{\cplxj}{\jmath}

\safemath{\dict}{\matD}
\safemath{\inputdim}{N}		
\safemath{\outputdim}{M}		
\safemath{\sparsity}{S}	
\safemath{\inputdimA}{{N_a}}	
\safemath{\inputdimB}{{N_b}}	
\safemath{\elemA}{{n_a}}	
\safemath{\elemB}{{n_b}}	
\safemath{\resA}{\matR_a}	
\safemath{\resB}{\matR_b}	
\safemath{\subD}{\matS} 
\safemath{\subA}{\matS_a} 
\safemath{\subB}{\matS_b} 
\safemath{\dicta}{\matA} 	
\safemath{\dictb}{\matB} 	
\safemath{\hollowS}{H}
\safemath{\hollowA}{H_a}
\safemath{\hollowB}{H_b}
\safemath{\cross}{Z}
\safemath{\coh}{\mu_d}			
\safemath{\coha}{\mu_a}			
\safemath{\cohb}{\mu_b}			
\safemath{\mubs}{\nu}	
\safemath{\cohm}{\mu_m} 
\safemath{\dictset}{\setD}	
\safemath{\dictsetp}{\dictset(\coh,\coha,\cohb)}	
\safemath{\dictsetgen}{\dictset_\text{gen}}
\safemath{\dictsetgenp}{\dictsetgen(\coh)}
\safemath{\dictsetonb}{\dictset_\text{onb}}
\safemath{\dictsetonbp}{\dictsetonb(\coh)}

\safemath{\leftside}{U}
\safemath{\rightsideA}{R_a}
\safemath{\rightsideB}{R_b}

\safemath{\indexS}{\setI_S} 

\safemath{\na}{n_a}			
\safemath{\nb}{n_b}			
\safemath{\coeffa}{p_i}	
\safemath{\coeffb}{q_j}	
\safemath{\seta}{\setP}		
\safemath{\setb}{\setQ}     
\safemath{\setw}{\setW}	
\safemath{\setz}{\setZ}	
\safemath{\cola}{\veca}		
\safemath{\colb}{\vecb}		
\safemath{\cold}{\vecd}		
\safemath{\inputvec}{\vecx} 	
\safemath{\error}{\vece}	
\safemath{\noiseout}{\vecz} 	
\safemath{\inputvecel}{x}
\safemath{\inputveca}{\vecx_a}
\safemath{\inputvecb}{\vecx_b}
\safemath{\outputvec}{\vecy}	
\safemath{\lambdamin}{\lambda_{\mathrm{min}}}

\safemath{\elltwo}{\ell_2}
\safemath{\ellone}{\ell_1}
\safemath{\ellzero}{\ell_0}
\safemath{\ellinf}{\ell_\infty}
\safemath{\ellinftilde}{\ell_{\widetilde\infty}}
\safemath{\licard}{Z(\coh,\coha,\cohb)}
\safemath{\xsol}{\hat{x}}
\safemath{\xbord}{x_b}		
\safemath{\xstat}{x_s}		
\safemath{\xstatLone}{\tilde{x}_s}
\safemath{\order}{\mathcal{O}} 
\safemath{\scales}{\Theta} 
\safemath{\ones}{\mathbf{1}} 
\safemath{\zeroes}{\mathbf{0}} 
\safemath{\thlone}{\kappa(\coh,\cohb)} 
\safemath{\constoneA}{\delta} 
\safemath{\constoneB}{\epsilon} 
\safemath{\nlarge}{L}				   
\safemath{\sumlarge}{S_\nlarge}
\safemath{\maxlarger}{P_\nlarge}	   
\safemath{\Pzero}{\textrm{P0}}	
\safemath{\Pone}{\textrm{P1}}
\safemath{\vecfir}{\vecw}			 
\safemath{\vecsec}{\vecz}
\safemath{\elvecfir}{w}              
\safemath{\elvecsec}{z}				 
\safemath{\nlargefir}{n}
\safemath{\normout}{\gamma}
\safemath{\auxfun}{h}
\safemath{\supp}{\textrm{supp}}

\safemath{\indexa}{\ell}
\safemath{\indexb}{r}
\safemath{\indexc}{i}
\safemath{\indexd}{j}

\safemath{\project}{P}

\usepackage{color}

\allowdisplaybreaks 

\safemath{\LAMA}{\textrm{LAMA}}
\safemath{\MRT}{\textrm{MRT}}
\safemath{\betamax}{\beta^\text{max}_\setO}
\safemath{\betamaxno}{\beta^\text{max}}
\safemath{\betamin}{\beta^\text{min}_\setO}
\safemath{\betaminno}{\beta^\text{min}}

\safemath{\Nomin}{\No^\textnormal{min}(\beta)}
\safemath{\Nominnobeta}{\No^\text{min}}
\safemath{\Nomax}{\No^\textnormal{max}(\beta)}
\safemath{\Nomaxnobeta}{\No^\textnormal{max}}
\safemath{\EX}{E_\textnormal{x}}
\safemath{\EXP}{\EX^\textnormal{p}}
\safemath{\Eo}{E_0}

\safemath{\tmax}{{t_\textnormal{max}}}
\safemath{\MAP}{\textrm{MAP}}
\safemath{\IO}{\textrm{IO}}
\safemath{\JO}{\textrm{JO}}
\safemath{\Nopost}{N_{0}^\textnormal{post}}
\safemath{\MT}{U}
\safemath{\MR}{B}
\safemath{\Tran}{\textnormal{T}}
\safemath{\Herm}{\textnormal{H}}
\safemath{\row}{\textnormal{r}}
\safemath{\col}{\textnormal{c}}

\safemath{\NT}{N_\textnormal{T}}
\safemath{\DSNR}{\delta \textnormal{SNR}}
\safemath{\betaMOR}{\beta^{\star}}

\begin{document}
	
\title{Reducing the Complexity of Fingerprinting-Based Positioning using Locality-Sensitive Hashing}

\author{\IEEEauthorblockN{Larry Tang, Ramina Ghods, and Christoph Studer} \\[-0.1cm]
\IEEEauthorblockA{\textit{School of Electrical and Computer Engineering, Cornell University, Ithaca, NY} \\
\textit{email: \{lwt29,\,rg548,\,studer\}@cornell.edu; website: {vip.ece.cornell.edu}}}
\thanks{The work of LT was supported in part by the Cornell University Engineering Learning Initiatives (ELI). 
The of RG and CS was supported in part by the semiconductor research corporation (SRC) nCORE task 2758.004, by Xilinx Inc., and by the US National Science Foundation under grants ECCS-1408006, CCF-1535897,  CCF-1652065, CNS-1717559, and ECCS-1824379. The authors thank E.~Lei and O.~Casta\~neda for their help with CSI generation.}
%
}

\maketitle
\begin{abstract}
Localization of wireless transmitters based on channel state information (CSI) fingerprinting finds widespread use in indoor as well as outdoor scenarios. 
Fingerprinting localization first builds a database containing CSI with measured location information. One then searches for the most similar CSI in this database to approximate the position of wireless transmitters.
In this paper, we investigate the efficacy of locality-sensitive hashing (LSH) to reduce the complexity of the nearest neighbor-search (NNS) required by conventional fingerprinting localization systems.
More specifically, we propose a low-complexity and memory efficient LSH function based on the sum-to-one (STOne) transform and use approximate hash matches.
We evaluate the accuracy and complexity (in terms of the number of searches and storage requirements) of our approach for line-of-sight (LoS) and non-LoS channels, and we show that LSH enables low-complexity fingerprinting localization with comparable accuracy to methods relying on exact NNS or deep neural networks. 
\end{abstract}


\section{Introduction}
Localization of wireless transmitters in indoor and outdoor scenarios has received growing interest over the decade. 
Conventional methods for outdoor localization are mainly based on triangulation or trilateration methods, which map time-of-flight (ToF), angle-of-arrival, or received signal strength (RSS) features to a specific location using geometric models~\cite{GG05,liu_survey_2007,KGKR14,garcia2017direct}. 
All of these approaches rely on line-of-sight (LoS) connectivity to multiple basestations, access points, or satellites. A popular instance of such methods are global navigation-satellite systems (GNSS) that combine ToF measurements with geometric and physical models. 
In situations that lack LoS connectivity (which is the case for indoor scenarios) or systems where GNSS is unavailable (which is the case for ultra-low-power sensors),  alternative positioning methods are required.

\subsection{Localization using CSI-Fingerprinting}
{Location fingerprinting} is a prominent method to enable wireless positioning in challenging propagation environments, such as non-LoS scenarios or channels with complex multi-path propagation; see, e.g.,~\cite{bahl_radar:_2000,kaemarungsi2004modeling,kaemarungsi2004properties,liu_survey_2007,bshara2010fingerprinting} and the references therein.
The principle of location fingerprinting is as follows. 
In a first phase, one tabulates a large number of measured channel-state information (CSI), which includes RSS~\cite{kaemarungsi2004modeling,kaemarungsi2004properties} or other measured channel features in the time or frequency domain~\cite{ibrahim2012cellsense,wang2015deepfi,wu_csi-based_2013} or at multiple receive antennas~\cite{chapre2015csi}, with associated location information in a given area---the resulting CSI-fingerprint/location tuples are stored in a database. 
In a second phase, a new CSI-fingerprint is extracted from a wireless transmitter and the most similar CSI fingerprints in the database are retrieved. The stored locations associated with the nearest fingerprints are then used to generate an estimate of the transmitter's location. 
While a  nearest neighbor search (NNS) in the CSI-fingerprint database can provide a simple (often accurate) estimate of the transmitter's location\footnote{More sophisticated methods, such as neural networks~\cite{smailagic2000determining} or Bayesian  methods~\cite{castro2001probabilistic} can be used to improve the nearest neighbor search estimate.}, the complexity of NNS in large fingerprinting databases can quickly become a bottleneck. 
More recently, localization approaches based on deep  neural networks have been proposed in~\cite{wang2016csi,vieira2017deep,arnold2018deep,lei2019siamese,studer2018channel}. Such methods avoid a NNS and directly map measured CSI-fingerprints to location---learning these networks, however, typically requires a large number of CSI measurements 
at fine resolution in space (often of the order of a few wavelengths). Furthermore, the large number of network parameters can easily be of the same order as the size of the fingerprint database.  

\subsection{Contributions}
In this paper, we reduce the complexity of the NNS step in traditional fingerprinting localization via locality-sensitive hashing (LSH)~\cite{gionis1999similarity,andoni_near-optimal_2008,charikar2002similarity,har2012approximate}. 
We design a computationally efficient LSH function that builds upon a randomized sum-to-one (STOne) transform~\cite{goldstein_stone_2015}, and we use approximate hash matches that reduce the complexity of the NNS and the size of the hash tables. 
We evaluate the accuracy and complexity (in terms of the number of searches and storage requirements) of our method for LoS and non-LoS channels in a massive multiuser (MU) multiple-input multiple-output (MIMO) wireless system. 
Finally, we compare the accuracy, complexity, and storage requirements of our approach to that of recent neural network-based localization methods proposed in~\cite{wang2016csi,vieira2017deep,arnold2018deep,lei2019siamese}.

\section{Fingerprinting Localization via LSH}
We now introduce the principles of fingerprinting localization and summarize the basics of LSH. We then detail our approach for low-complexity LSH with approximate hash lookups. 

\subsection{Basics of Fingerprinting Localization }
Fingerprinting localization proceeds in two phases. In the first phase,  CSI-fingerprints $\{\bmf_n\}_{n=1}^N=\setF$ and associated positions $\{\bmx_n\}_{n=1}^N=\setX$ are measured in a given area and  stored in a database.  
Here, the vector $\bmf_n\in\reals^D$ corresponds to a $D$-dimensional CSI-fingerprint at position $\bmx_n\in\reals^d$, where $D$ is typically high-dimensional and $d$ is either two or three dimensions.
Depending on the application, CSI-fingerprints can represent RSSs acquired at multiple receivers, power-delay profiles, angle-of-arrival, and many others; see~\cite{liu_survey_2007} for a survey.

In the second phase, an estimate for the location~$\bmx_{n'}$ of a new transmitting device with index $n'$ is generated. To this end, a CSI-fingerprint $\bmf_{n'}$ is extracted. Then, the indices associated with the $K$ most similar fingerprints in $\{\bmf_n\}_{n=1}^N$ are identified 
\begin{align*}
\setN_K = \{n : \|\bmf_n-\bmf_{n'}\|<r,\bmf_n\in\setF,n=1,\ldots,N\},
\end{align*}
where $r>0$ ensures that $|\setN_K|=K$. One then approximates the location of transmitter $n'$ from the set of similar locations $\setX_K=\{\bmx_n\}_{n\in\setN_K}$. 
If $K=1$, then one can simply pick the location associated with the nearest CSI fingerprint; more sophisticated approaches are  discussed in~\cite{smailagic2000determining,castro2001probabilistic}.

\subsection{Locality-Sensitive Hashing (LSH)}
\label{sec:LSH}

Finding the $K$ nearest neighbors in a large dataset containing high-dimensional vectors suffers from the curse of dimensionality; this implies that carrying out an exhaustive search inevitably results in high complexity~\cite{andoni_near-optimal_2008}.  
Locality-sensitive hashing (LSH) is a powerful method to perform an approximate nearest neighbor search in large datasets \cite{gionis1999similarity,andoni_near-optimal_2008,charikar2002similarity,har2012approximate}. 
The principle behind LSH is to construct locality-sensitive hash functions for which similar datapoints have matching hash values and dissimilar datapoints have mismatched hash values.
Mathematically, LSH relies on functions $h:\reals^D \to \setS$ 
for which the  hash collision probability $\Pr[h(\textbf{p}) = h(\textbf{q})]=P_1$ of any two datapoints $\bmp,\bmq\in\reals^D$ is large if $\|\bmp-\bmq\|\leq R$  and $\Pr[h(\textbf{p}) = h(\textbf{q})]=P_2$ is small if $\|\bmp-\bmq\| \geq c R$ with~$R$ being an application-dependent radius and $c>1$ a constant; for LSH, one is particularly interested in the case $P_1 \gg P_2$. The finite-cardinality set~$\setS$ contains so-called buckets, which are nothing but unique hash values.

Approximate NNS via LSH is carried out in two phases. In the first phase, one computes hash values for all points in the dataset, i.e.,  $\{h(\bmf_n)\}_{n=1}^N$. In the second phase, for a given query point $\bmf_{n'}$, one computes $h(\bmf_{n'})$ and compares the resulting hash value to those in the dataset. One can then compare the true, high-dimensional CSI feature distance   $\|\bmf_m-\bmf_{n'}\|$ associated to \emph{only} those indices $m\in\{n:h(\bmf_n)=h(\bmf_{n'}),n=1,\ldots,N\}$ for which there was a collision.
By using a set of $T$ distinct and carefully crafted LSH functions (instead of just one LSH function), one can improve the likelihood to find at least one nearest neighbor while often significantly reducing the complexity of approximate NNS, even for very large datasets.

\subsection{Fast LSH via the Randomized STOne Transform}
The  literature describes a number of ways to construct LSH functions with the desired properties~\cite{andoni_near-optimal_2008}.  
In what follows, we are interested in LSH functions that can be computed at low complexity and require low memory footprint---our goal is to reduce the complexity of LSH-based location fingerprinting. 

Our approach builds upon prominent LSH functions for normalized datapoints~\cite{andoni_practical_2015} and uses ideas from \cite{kapralov_how_2016,goldstein_stone_2015} to lower the complexity. 
In \cite{andoni_practical_2015}, the authors introduce a family of hash functions called cross polytope LSH for applications in which data points are distinguished by angular distance.
For cross polytope LSH, one hashes a point $\bmx \in \mathbb{R}^D$ that lies on the unit sphere by computing $\bmy = \bA\bmx/\|\bA\bmx\|$, where $\bA \in \mathbb{R}^{D \times D}$ is a random matrix with i.i.d.\ Gaussian entries. The closest standard basis vector of $\mathbb{R}^D$ to $\bmy$ is then used as the hash of $\bmx$. 
To make cross polytope hash functions more practical, reference~\cite{andoni_practical_2015} suggests the use of pseudo-random rotations.  Concretely, instead of multiplying an input vector $\bmx$ by a random Gaussian matrix, one can ``fake multiply'' it with $\bH\bD$, where $\bH$ is a Hadamard matrix (which has a fast transform) and $\bD$ is a diagonal matrix with a fixed but random sequence of antipodal entries \cite{kapralov_how_2016}.

In our application, we select a diagonal matrix $\bD\in\reals^{D\times D}$ in which the diagonal entries are pseudo-random $\pm1$ values. 
We also select a pseudo-random index set $\Omega\subset\{1,2,\ldots,D\}$, where $|\Omega|=L$ is the length of the hash values.
For $\bH$, we use the sum-to-one (STOne) transform matrix $\bH\in\reals^{D\times D}$, which is a Hadamard transform that has (i) $D\log(D)$ complexity and (ii) multi-scale properties~\cite{goldstein_stone_2015}.
The STOne transform matrix can be constructed using a $4\times4$ stencil matrix defined as~\cite{goldstein_stone_2015} 
\begin{align}
S_4 = \dfrac{1}{2} \begin{bmatrix} -1 & 1 & 1& 1 \\  1 &-1 & 1& 1 \\ 1 & 1 & -1& 1 \\ 1 & 1 & 1& -1 \\\end{bmatrix}\!.
\end{align}
One then forms a larger STOne transform matrix by computing the Kronecker product of two (or more) stencil matrices, e.g., $S_{16} = S_4\otimes S_4$ where $\otimes$ is the Kronecker product.

With these ingredients, we can design the following LSH function: $h(\bmf) = [\sign(\bH\bD\bmf)]_\Omega$. Here, the operator $[\cdot]_\Omega$ extracts the vector whose entries are associated to the indices in $\Omega$; $\sign(\cdot)$ operates element-wise on vectors.
For this LSH function, each hash value (also called bucket) is in the set $\setS=\{-1,+1\}^L$.
Since $\tilde\bmf=\bD\bmf$ can be computed in linear time and the STOne transform $\bH\tilde\bmf$ in $D\log(D)$ time~\cite{goldstein_stone_2015}, computing hash values $h(\bmf)$ is efficient and requires a minimum amount of storage (only for diagonal entries of $\bD$ and the subset~$\Omega$). 
	
To design multiple hash functions and tables, we simply generate~$T$ pseudo-random subsets $\Omega_t\subset\{1,\ldots,D\}$, $t=1,\ldots,T$. We then compute~$T$ hash functions as $h_t(\bmf) = [\sign(\bH\bD\bmf)]_{\Omega_t}$, $t=1,\ldots,T$, which requires the computation of $\sign(\bH\bD\bmf)$ only once (instead of computing $T$ Hadamard transforms). 

\begin{figure*}[tp]
\centering
\subfigure[Fraction compared; LoS]{\includegraphics[width=0.33\textwidth]{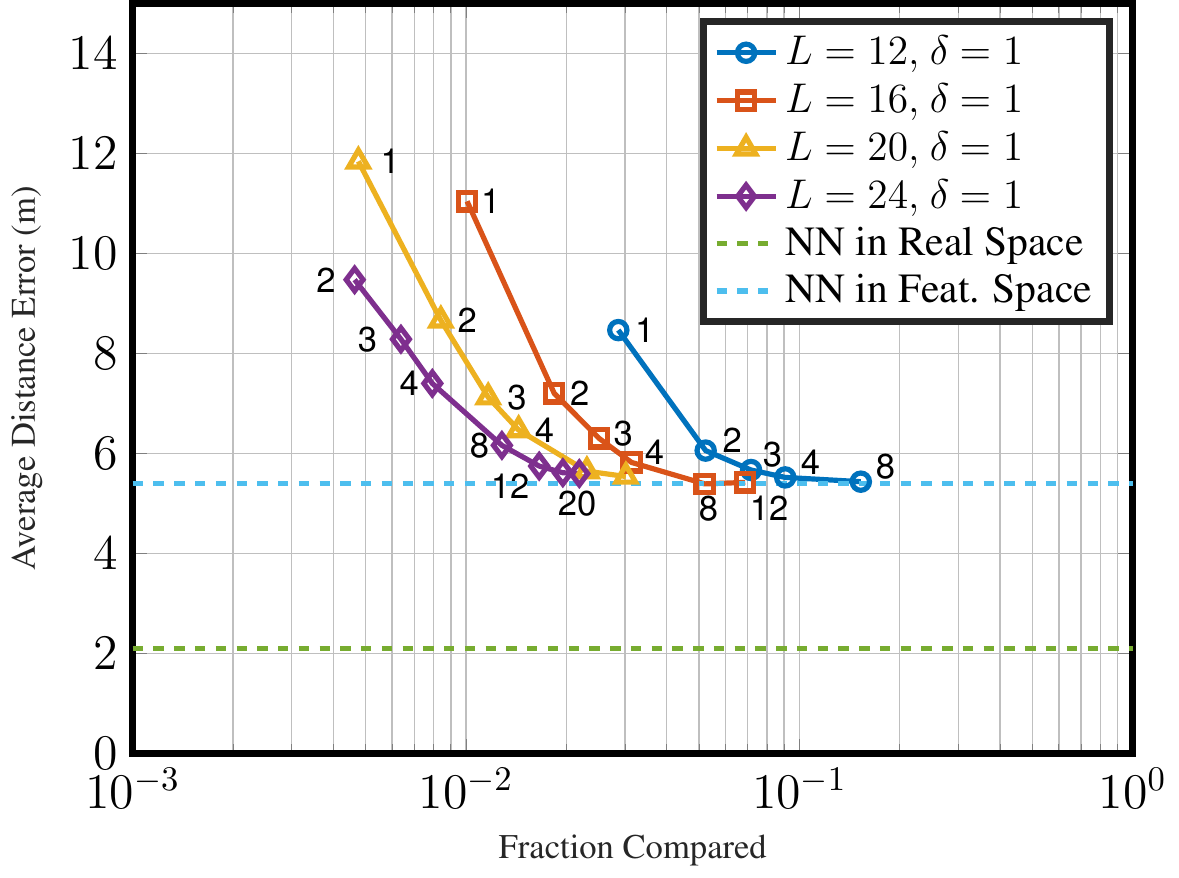}}
\subfigure[Memory area; LoS]{\includegraphics[width=0.33\textwidth]{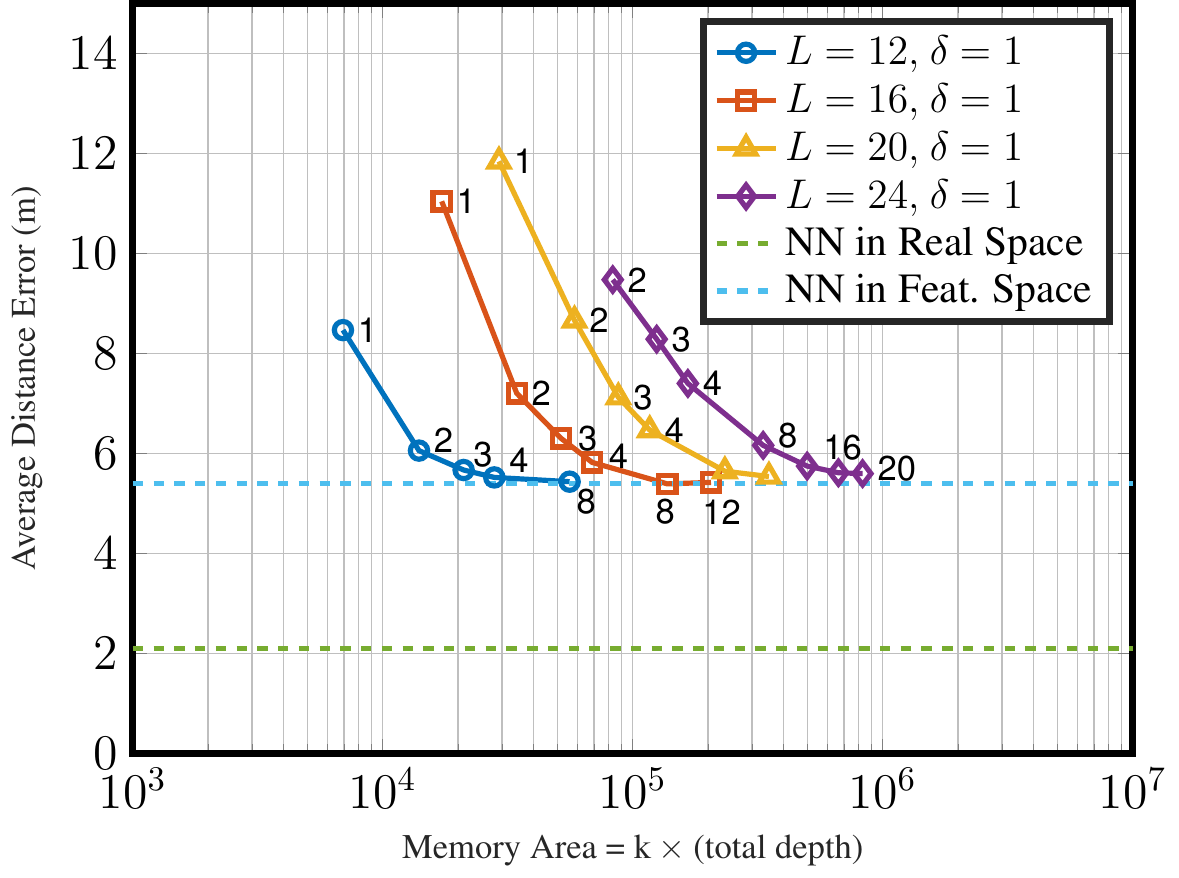}}
\subfigure[Total complexity; LoS]{\includegraphics[width=0.32\textwidth]{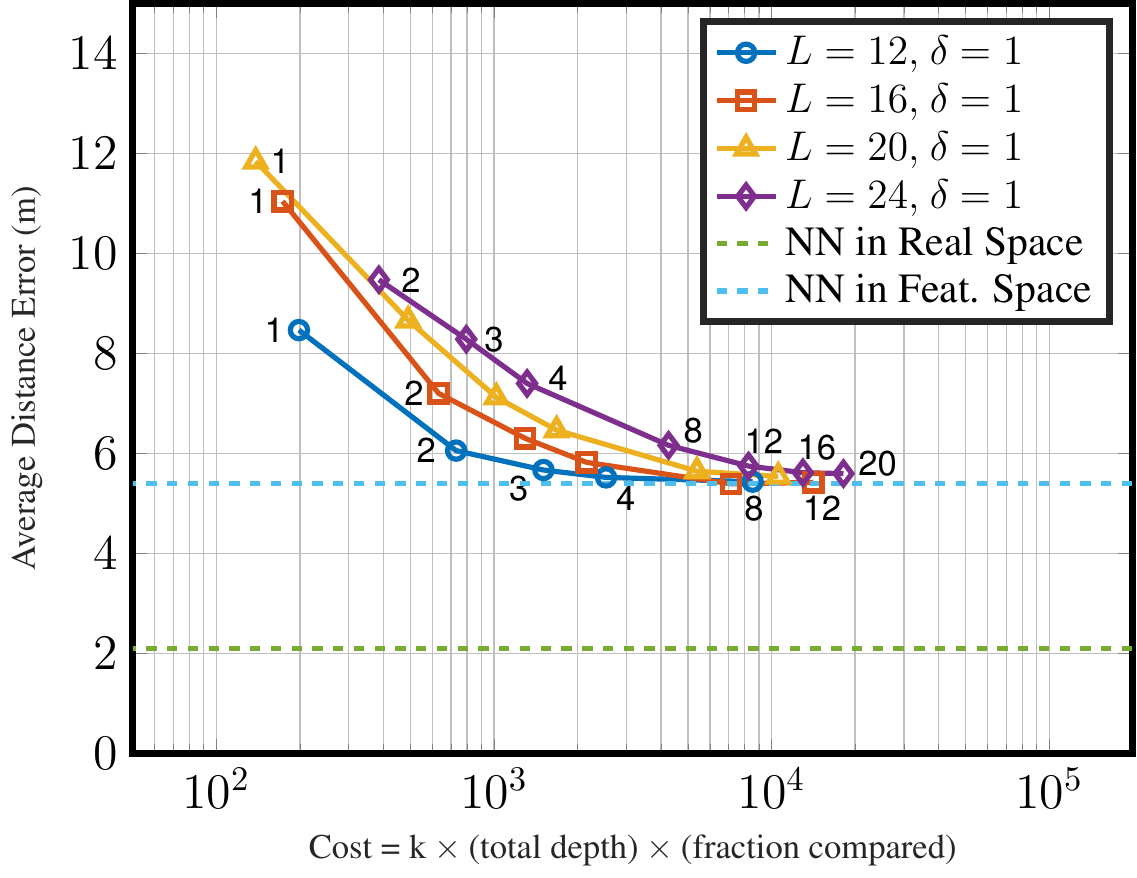}}
\subfigure[Fraction compared; non-LoS]{\includegraphics[width=0.33\textwidth]{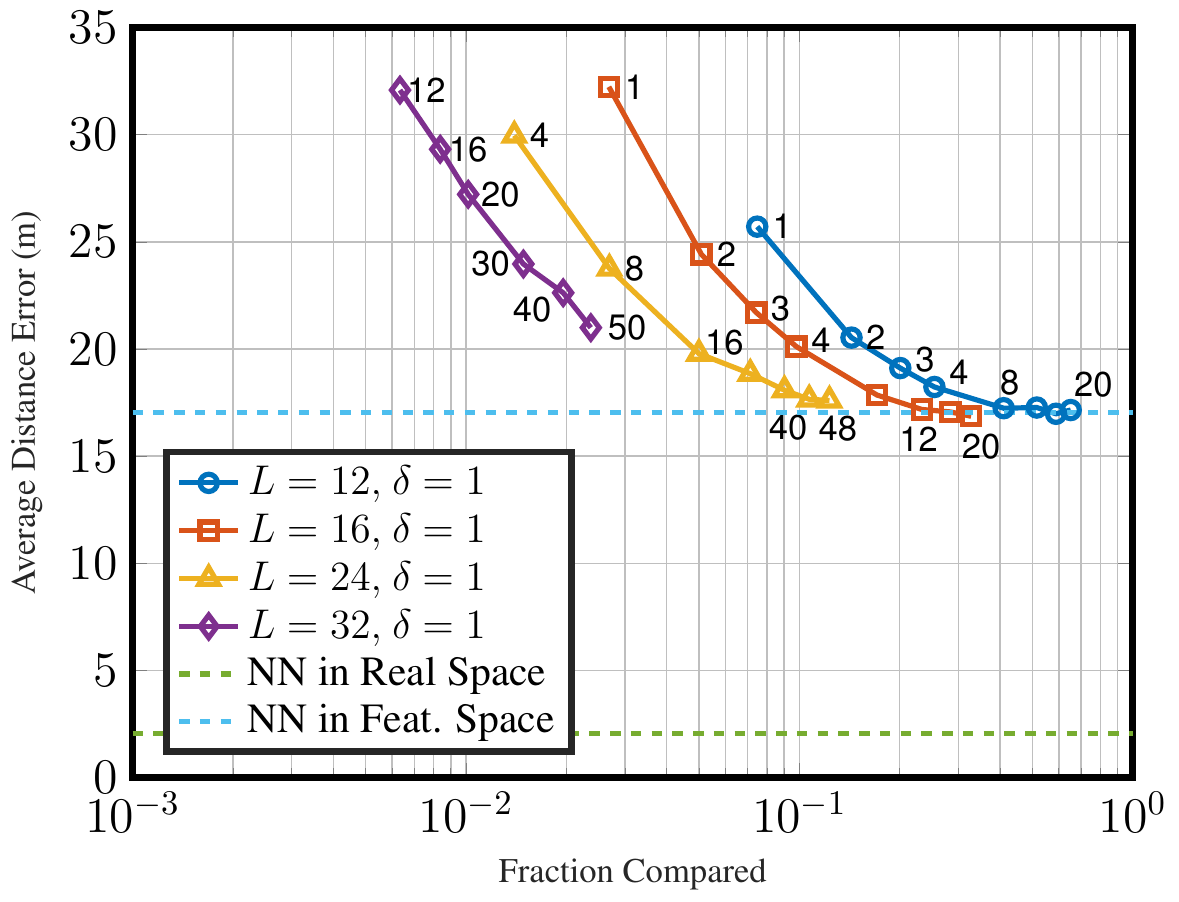}}
\subfigure[Memory area; non-LoS]{\includegraphics[width=0.33\textwidth]{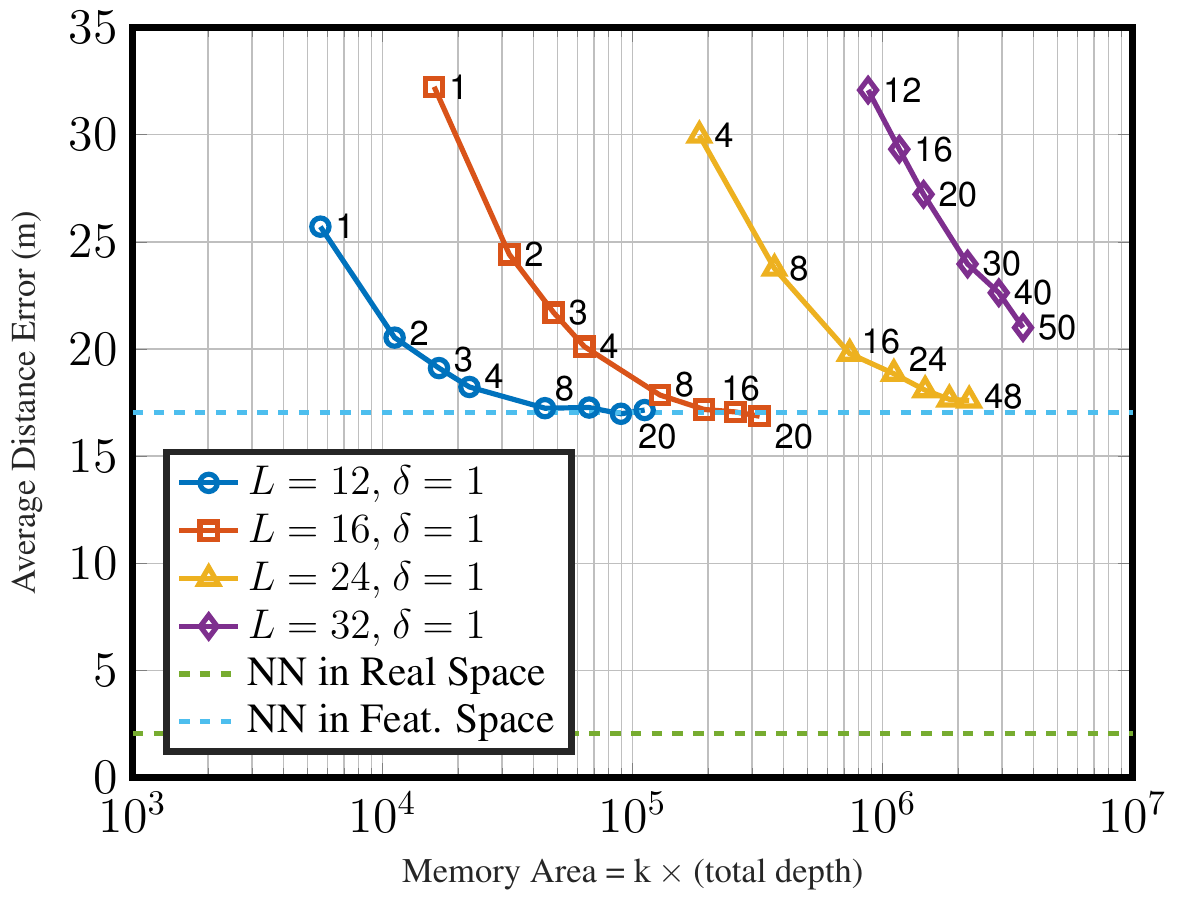}}
\subfigure[Total complexity; non-LoS]{\includegraphics[width=0.32\textwidth]{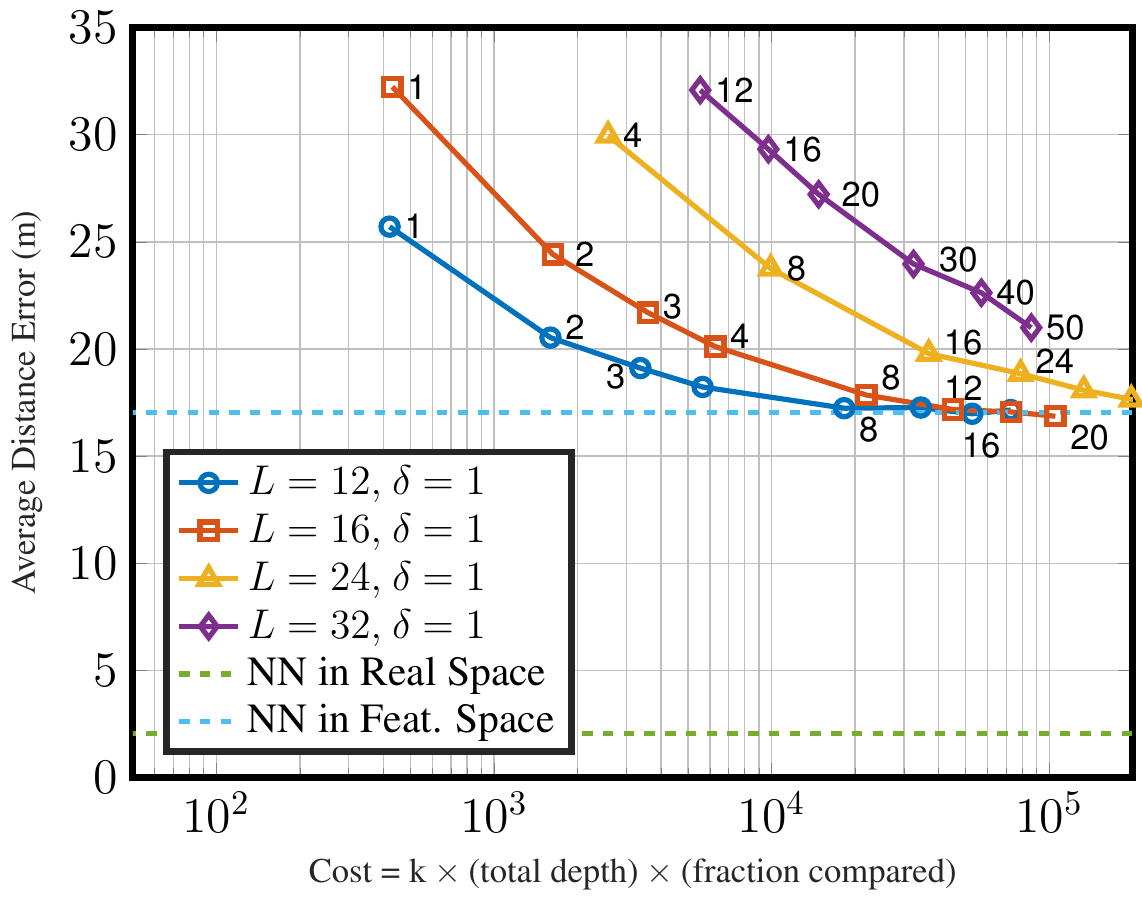}}
\caption{Accuracy of LSH-based location fingerprinting for LoS and non-LoS massive MU-MIMO scenarios. "NN in feature space" are the nearest neighbors found through an exhaustive search and "NN in real space" are the true nearest neighbors in real space.}
\label{fig:simulation}
\vspace{-0.13cm}
\end{figure*}

\subsection{LSH with Approximate Matches}
Our final ingredient addresses the storage requirements of LSH-based location fingerprinting. 
To this end, we declare a match whenever the Hamming distance between $h(\bmp)$ and $h(\bmq)$ is within a given threshold $\delta \geq0$ (instead of declaring a match only if the two hash values are equal). Note that for $\delta=0$, we perform classical LSH. The key advantage of approximate matches (i.e., $\delta>0$) is that fewer hash tables are sufficient to achieve the same performance (in terms of finding the $K$ nearest neighbors) as conventional LSH; see the next section for more details.  Furthermore, recent advances in processing-in-memory (PIM) enables one to quickly identify approximate hash matches in a hardware efficient manner~\cite{CBGS19}.

\section{Results and Comparisons}
\label{sec:results}
We now show results of the proposed low-complexity LSH-based fingerprinting localization method. We provide a detailed accuracy/complexity trade-off analysis and a comparison to recent neural-network localization approaches~\cite{wang2016csi,vieira2017deep,arnold2018deep}. 
Our main goals are as follows: (i) reduce the complexity of searching similar CSI fingerprints in large databases and (ii) reduce the complexity at minimal storage overhead. 

\subsection{Simulated Scenario}
To evaluate the efficacy of our approach, we consider a massive MU-MIMO-OFDM localization scenario in LoS and non-LoS scenarios with a single basestation containing $32$ antennas operating at $2.68$\,GHz with a bandwidth of $20$\,MHz and localizing $2000$ transmitters distributed uniformly at random in an area of $40,000$\,m$^2$; the noisy channel vectors are generated using channel models from~\cite{QuaDRiGa}. The CSI features are $D=256$ dimensional ($32$ antennas and $8$ maximally-spaced subcarriers) and correspond to the absolute value of beamspace/delay domain channel vectors as in~\cite{vieira2017deep,studer2018channel}. We use the $K=2$ nearest neighbors and average their location; we also use approximate matches with a Hamming distance of $\delta=1$. Both of these algorithm choices yield consistently good results. Below, we provide simulation results that confirm this claim. 
To assess the complexity and storage requirements, we use the following metrics: (i) the fraction of high-dimensional vectors that have been compared relative to an exhaustive search; (ii) the memory area indicating the size of the precomputed LSH tables; and (iii) the total complexity, which is the fraction of comparisons times the memory area.  

\subsection{Accuracy vs.\ Complexity vs.\ Storage Trade-offs}
Figure~\ref{fig:simulation} shows results for a LoS and a non-LoS scenario. As a reference, we include the performance of an exhaustive search (denoted by ``NN in feat. space'') and that of finding the true nearest neighbors in real space (denoted by ``NN in real space''). Each curve is parametrized by the number of hash functions $T$.  
We see in \fref{fig:simulation}(a)  that increasing the hash length $L$ can reduce the number of comparisons by 10$\times$, while achieving the same average distance error as an exhaustive search (indicated by the dashed ``NN in feat.~space'' lines).  
As shown in \fref{fig:simulation}(b), we see that increasing~$L$ increases the storage requirements for the hash values. 
This trade-off between reduced complexity but increased memory area is a result of the increase in the number of required hash tables but more sparsely populated hash buckets.
Similarly, we see that increasing the number of hash functions $T$ reduces the average distance error but increases the number of comparisons and the memory area. 
The total complexity is defined as fraction of comparisons times the memory area and is shown in \fref{fig:simulation}(c), which reveals that the best trade-off in terms of fraction of comparisons and memory area is achieved by relatively small hash values, i.e., with $L=12$ bits, and a few hash tables, i.e., with $T=4$ tables. 
Figures~\ref{fig:simulation}(d,e,f) show results for a non-LoS scenario. The trends are similar to that of the LoS case, which confirms that the proposed method is able to reduce the NNS complexity of LSH-based fingerprinting localization under different propagation conditions.

\begin{figure}[tp]
	\centering
	\includegraphics[width=0.7\columnwidth]{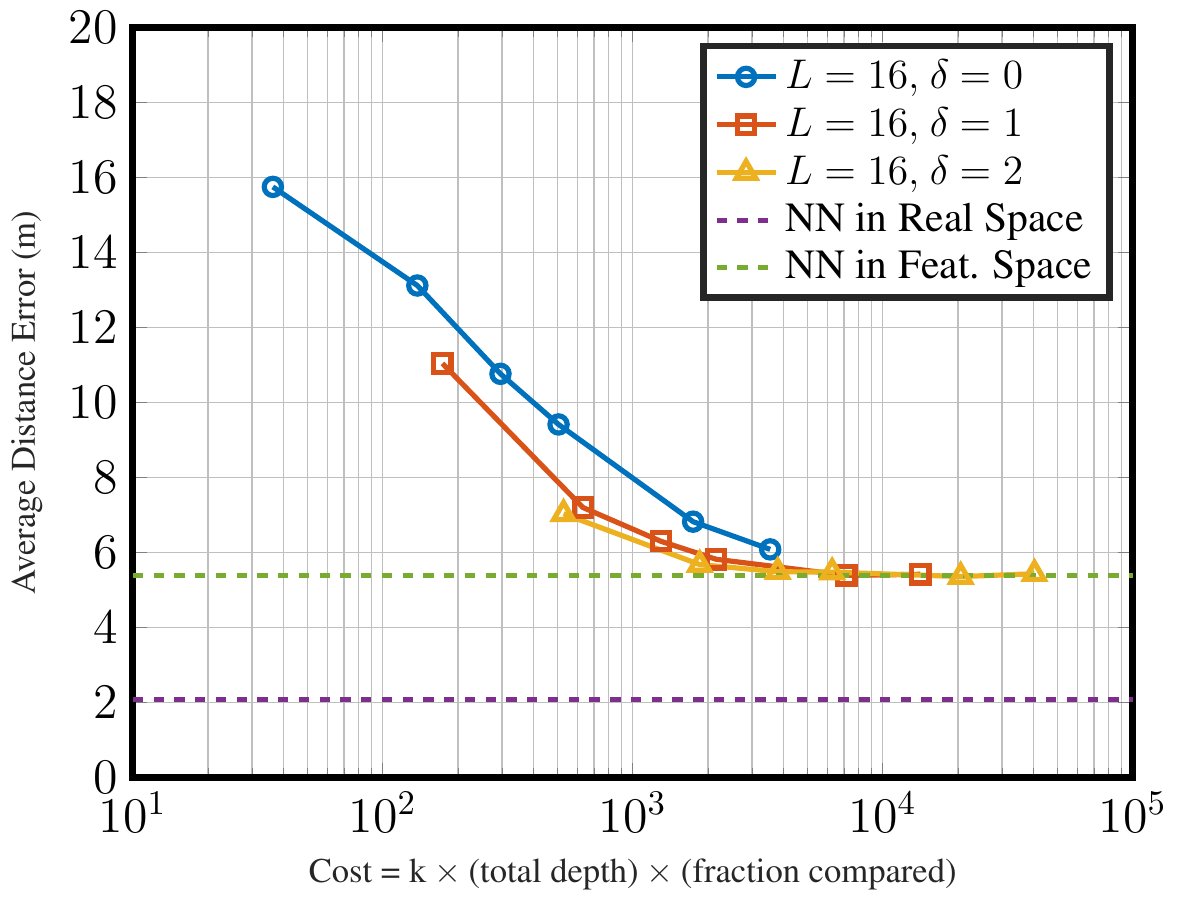}
	\caption{Total complexity of LSH-based location fingerprinting with approximate Hamming matches $\delta\in\{0,1,2\}$ for a LoS scenario.}
	\label{fig:delta}
\end{figure}

\begin{figure*}[tp]
	\centering
	\subfigure[Search complexity; LoS]{\includegraphics[width=0.34\textwidth]{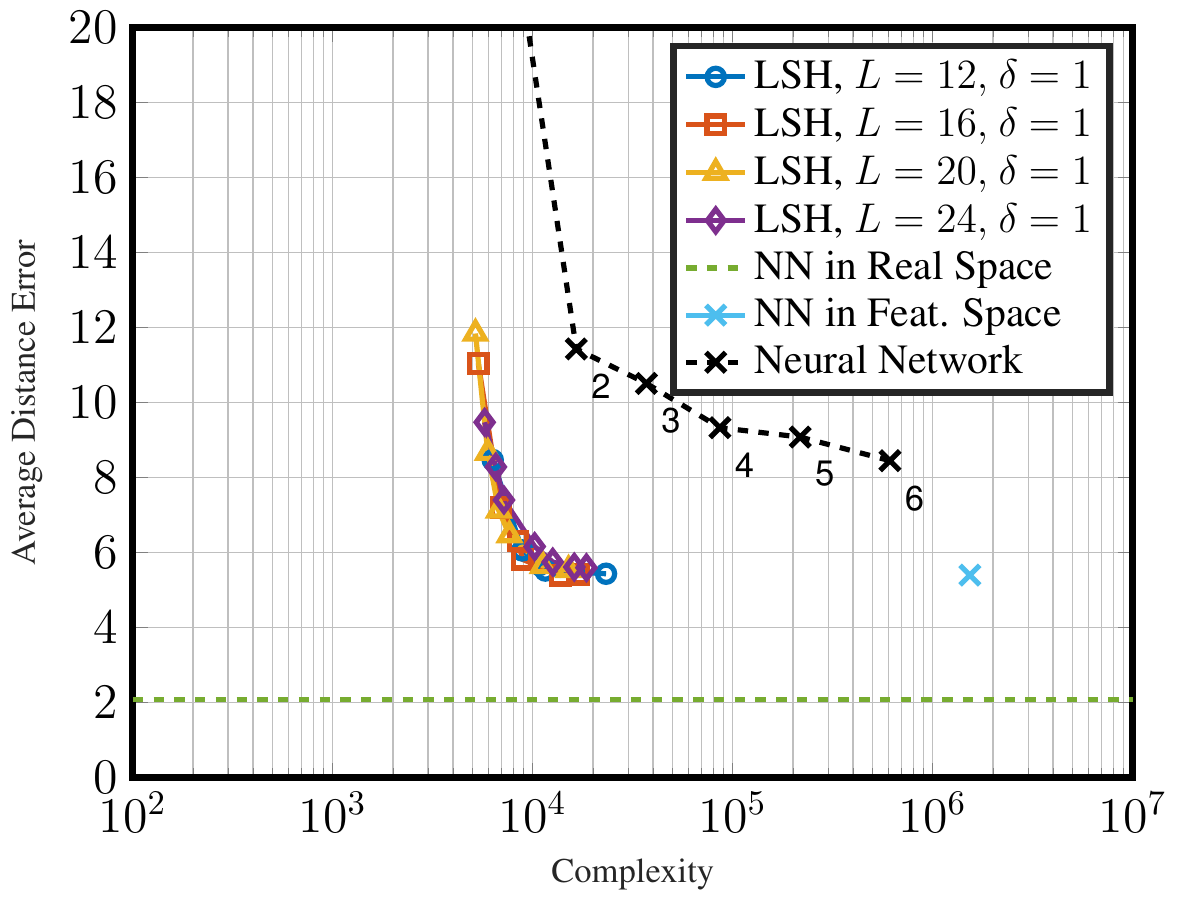}}
	\hspace{2cm}
	\subfigure[Memory area; LoS]{\includegraphics[width=0.34\textwidth]{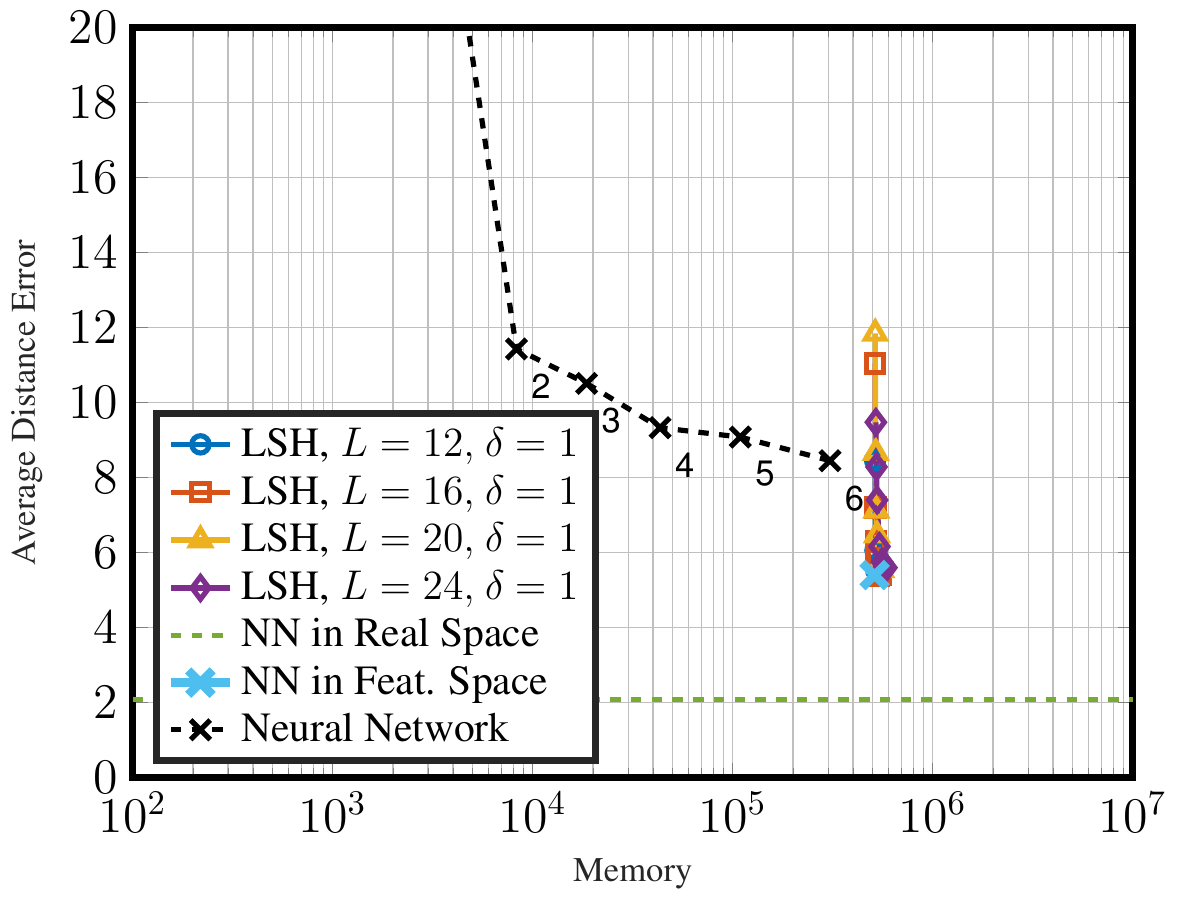}} \\
	\subfigure[Search complexity; non-LoS]{\includegraphics[width=0.34\textwidth]{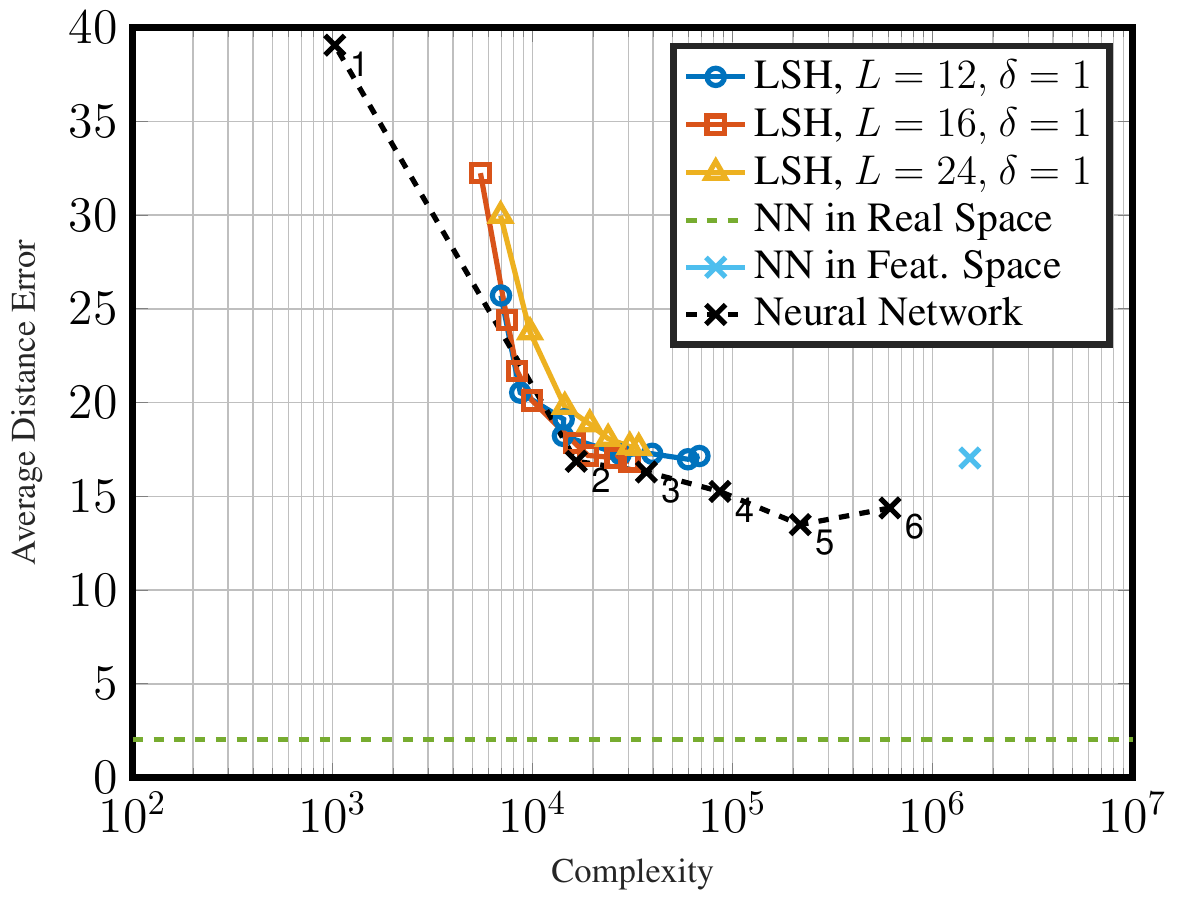}}
	\hspace{2cm}
	\subfigure[Memory area; non-LoS]{\includegraphics[width=0.34\textwidth]{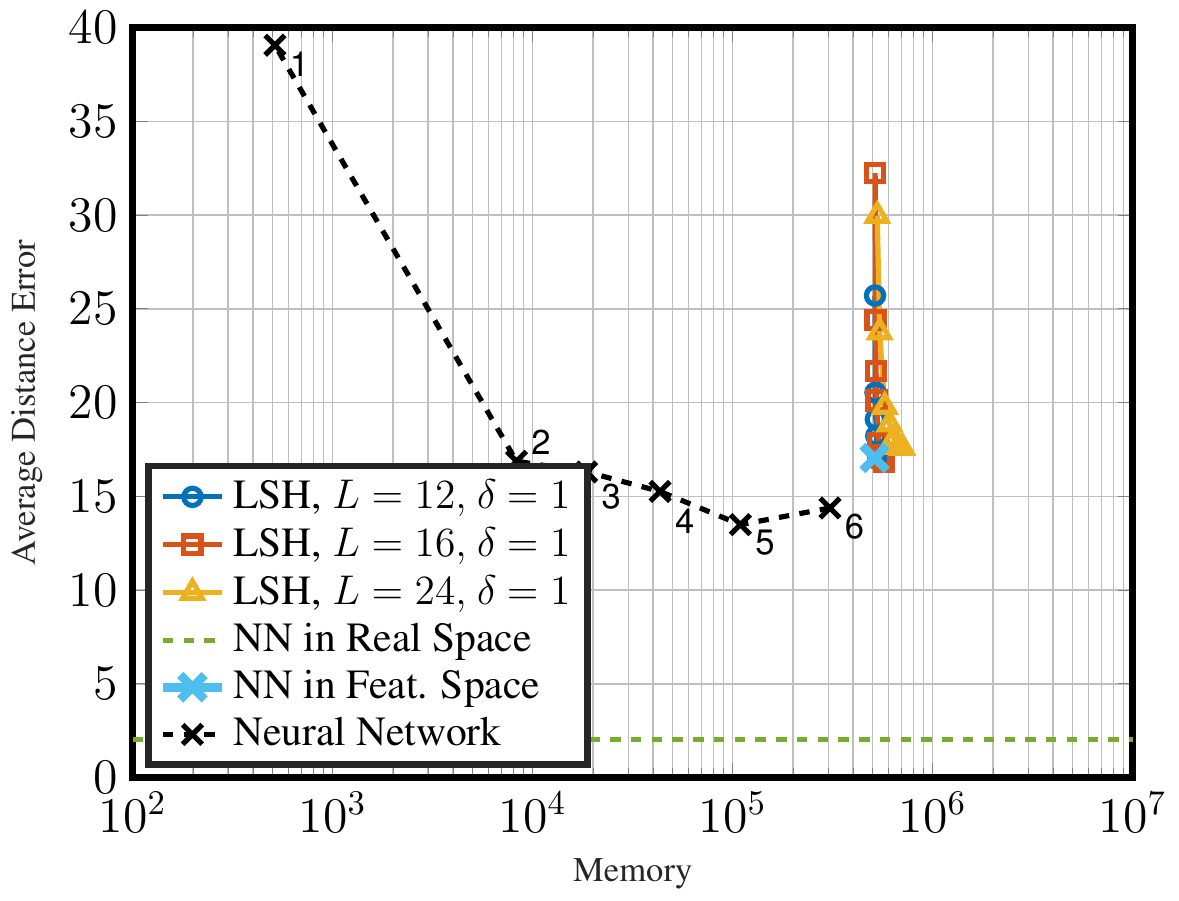}}	
	\caption{Comparison between LSH-based fingerprinting localization and neural network-based positioning. For LoS propagation conditions, LSH-based positioning reduces complexity while achieving superior accuracy; the memory requirements for neural network-based positioning is generally lower. For non-LoS propagation conditions, neural network-based positioning slightly outperforms LSH-based positioning in terms of accuracy, complexity, and memory. }
	\label{fig:neuralnetwork}
\end{figure*}

Figure~\ref{fig:delta} shows results for the LoS scenario where we compare different Hamming-distance thresholds $\delta$ to find exact or approximate matches.  We see that increasing $\delta$ from~$0$ to~$1$ reduces the average distance error for the same total costs as for $\delta=0$. Increasing $\delta$ to $2$  yields only a marginal improvement in terms of the average distance error, confirming that the $\delta=1$ threshold is a reasonable choice.
We would like to reiterate that approximate hash matches can be found efficiently using specialized PIM hardware, such as the one in~\cite{CBGS19}.

\subsection{Comparison to Deep Neural Networks}
We now compare the performance of our LSH-based approach to that of a deep neural network in terms of complexity and memory area.  We use fully connected neural networks (FCNNs) with different numbers of layers ($1$ to $6$) to compare accuracy, complexity, and memory area. The 6-layer FCNN has been used in \cite{lei2019siamese} and consists of $512$, $256$, $128$, $64$, $32$, and $2$ activations per layer; each layer uses rectified linear unit (ReLU) except the last layer, which uses a linear activation.
To obtain the FCNNs with $5$ to $1$ layers, we successively remove the layer with the largest number of nodes.  

To evaluate the complexity of LSH-based localization and FCNNs, we compute the total number of multiplications and additions required to find the approximate position of a single test point. In our LSH-based approach, we compute the complexity using the average number of comparisons needed for a given test point.  
The memory area for both approaches is the total number of stored values and parameters.  

Figures~\ref{fig:neuralnetwork}(a,b,c,d) show results for LoS and non-LoS scenarios. As a reference, we show the performance of an exhaustive search and that of finding true nearest neighbors in real space.  For the LoS scenario, Figs.~\ref{fig:neuralnetwork}(a,b) show that the LSH-based approach can achieve lower average distance error with lower complexity than the neural network based approach.  In terms of memory area,  we observe that the FCNN consistently performs better since it only needs to store weights and bias values, whereas the LSH-based approach requires storage for the entire fingerprint/location database so that comparisons can be made to points in CSI feature space.  For the non-LoS scenario, Figs.~\ref{fig:neuralnetwork}(c,d) show that the LSH-based approach no longer outperforms the neural network in terms of search complexity, i.e., FCNNs achieve slightly better average distance error with about the same complexity as the LSH-based approach.  We see similar trends in memory area as we did in the LoS scenario. 

\section{Conclusions}
\label{sec:conclusions}

We have demonstrated that the complexity of fingerprinting-based localization can be reduced significantly by using locality-sensitive hashing (LSH). 
To this end, we have proposed computationally efficient LSH functions that build upon the sum-to-one (STOne) transform and approximate hash matches.
Our results have shown that LSH-based fingerprinting is able to achieve the same accuracy as an exhaustive search over the CSI fingerprinting database for line-of-sight (LoS) and non-LoS scenarios, but at orders-of-magnitude lower complexity and storage requirements.
We have also demonstrated that LSH-based positioning is able to outperform neural network-based approaches under LoS propagation conditions. However, we have not observed any advantage of our approach for non-LoS propagation conditions. These observations imply that (i) the best approach for fingerprinting positioning depends strongly on the propagation conditions and (ii) neural networks are not necessarily the best approach for localization in terms of complexity and positioning accuracy.

\balance
\bibliographystyle{IEEEtran}
\bibliography{VIPabbrv,publishers,confs-jrnls,lsh}
\balance

\end{document}